\documentclass{vldb}
\usepackage{graphicx}
\usepackage{balance}  
\usepackage{subfigure}
\usepackage{amsmath}
\usepackage{wrapfig}
\usepackage{dsfont}
\usepackage{slashbox}
\usepackage[table]{xcolor}
\usepackage{color, colortbl}
\usepackage{algorithm}
\usepackage{algorithmic}

\usepackage{setspace}
\usepackage{mathrsfs}
\usepackage{cite}
\usepackage{enumerate}
\usepackage{amsfonts}
\usepackage{amsmath}
\usepackage{amssymb}
\usepackage{dsfont}
\usepackage{graphicx}

\newcommand{\Rset}{\mathop{\mathds{R}}}

\long\def\ftn[#1]#2{\begingroup%
\def\thefootnote{\fnsymbol{footnote}}\footnote[#1]{#2}\endgroup}

\definecolor{Gray}{gray}{0.9}

\begin{document}


\title{Rating through Voting: An Iterative Method for Robust Rating}



%
%
%
%

\numberofauthors{1} 

\author{
%
%
\alignauthor
Mohammad~Allahbakhsh, and Aleksandar~Ignjatovic\\
       \affaddr{The University of New South Wales}\\
        \affaddr{Sydney, NSW 2052, Australia}\\
       \email{\{mallahbakhsh,ignjat\}@cse.unsw.edu.au}
}

\maketitle

\begin{abstract}
In this paper we introduce an iterative voting algorithm and then use it to obtain a rating method which is very robust against collusion attacks as well as random and biased raters. Unlike the previous iterative methods, our method is not based on comparing submitted evaluations to an approximation of the final rating scores, and it entirely decouples credibility assessment of the cast evaluations from the ranking itself. The convergence of our algorithm relies on the existence of a fixed point of a continuous mapping which is also a stationary point of a constrained optimization objective. We have implemented and tested our rating method using both simulated data as well as real world data. In particular, we have applied our method to movie evaluations obtained from MovieLens and compared our results with IMDb and Rotten Tomatoes  movie rating sites. Not only are the ratings provided by our system very close to IMDb rating scores, but when we differ from the IMDb ratings, the direction of such differences is essentially always towards the ratings provided by the critics in Rotten Tomatoes. Our tests demonstrate high efficiency of our method, especially for very large online rating systems, for which trust management is both of the highest importance and one of the most challenging problems.
\end{abstract}

\section{Introduction}

Human computation is a new model of distributed computing~\cite{csproblemsolving,HCOmpsurvey} in which the computational power of machines is augmented by the cognitive power of human beings. The main benefit of such a model comes from the fact that many problems which are trivial to humans are still intractable for machines. Human computation has been employed to solve a wide variety of problems such as spam detection~\cite{spam}, question answering~\cite{so}, tagging photos~\cite{ESP}, as well as many others~\cite{cswww}.

E-Commerce is another area in which human computation has been widely used to assess quality of products as well as trustworthiness of people. Continuous growth of online commerce as well as of many other forms of online interactions, crucially depends on trust management. For example, in online markets, due to a huge number of potential sellers with varying reputation as well as a huge number of available products, sometimes of dubious quality, buyers rely on the feedback of other customers who have shared their experiences with the community to help them make their decisions. One common way of sharing such information and experiences is through \emph{Online Rating Systems}. In such systems, providers (either manufacturers or just sellers) advertise their products and customers evaluate them based on their experience of dealing with that particular product or vendor. Based on such individual evaluations received from customers, the system determines a rating score for every product, reflecting the overall quality of the product from customers' point of view. Yelp~\cite{yelp}, IMDb~\cite{imdb} and Amazon~\cite{amazon} are some of the popular online systems with rating facilities.

\begin{figure*}
\centering

\subfigure[Rating a product]{
\includegraphics[scale=0.5]{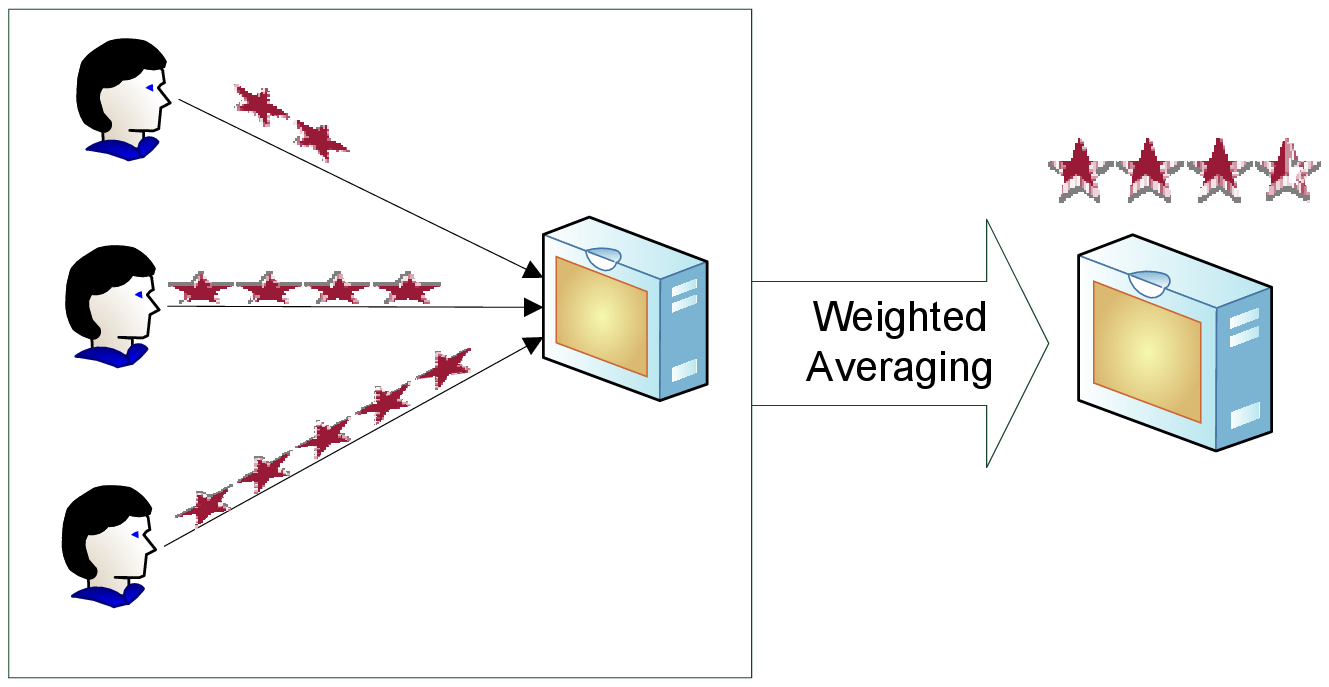}
\label{fig:r}
}\hspace{2em}
\subfigure[Corresponding Voting Scheme]{
\includegraphics[scale=0.6]{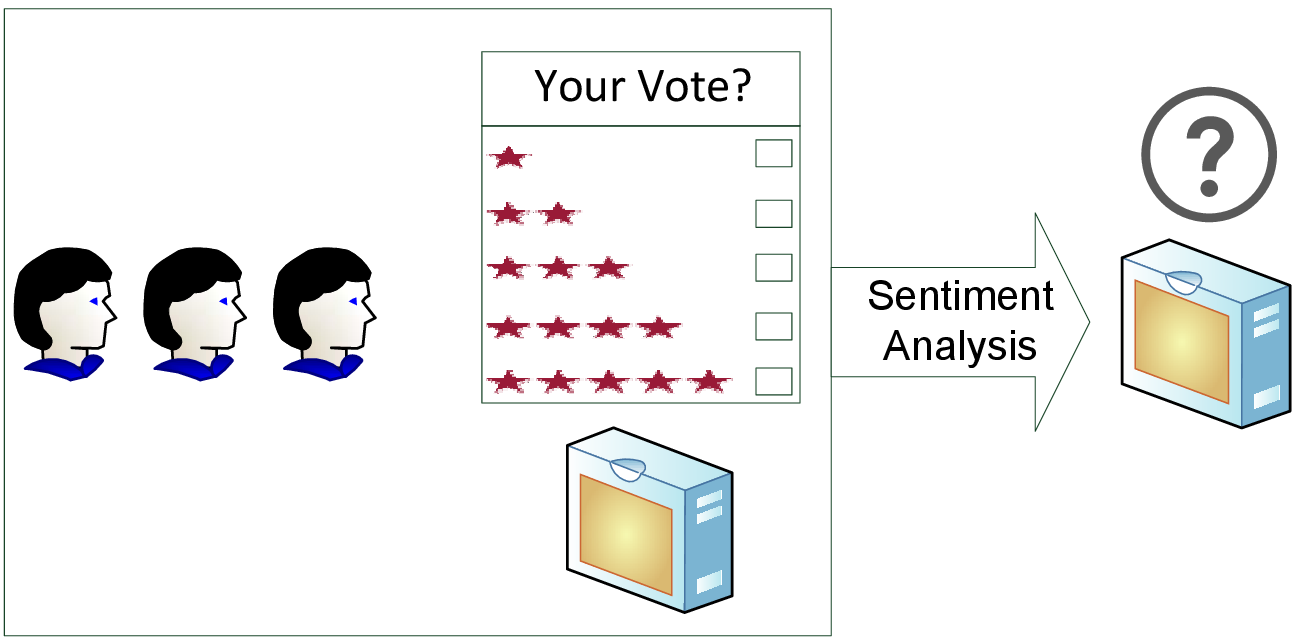}
\label{fig:v}
}
\label{fig:rtv}
\caption[]{Rating Through Voting - Method Overview}
\end{figure*}

An important issue in these systems is the trustworthiness of the cast feedback. Many pieces of evidence~\cite{ebayProblem,amazonproblem} show that users may try to manipulate the ratings of products by casting \emph{unfair evaluations}. Unfair evaluations are evaluations which are cast regardless of the quality of the product and usually are given based on personal vested interests of the users. For example, providers may try to submit supporting feedback to increase the rating of their products and consequently increase their revenue~\cite{amazonproblem}. The providers also may attack their competitors by giving low scores in their feedback on their competitor's products. Another study shows that some sellers in eBay boost their reputation unfairly by buying or selling feedback~\cite{ebayProblem}.

Unfair evaluations are broadly divided into two categories \cite{dishonest,reliable}: (i)~\emph{individual} and (ii)~\emph{collaborative}. Unlike the individual unfair evaluations, collaborative unfair evaluations, also called \emph{collusion}~\cite{IEEE2102Survey,reliable}, are cast by a group of users who try to manipulate the rating of a product collaboratively. Collusion, by its nature, is more sophisticated and harder to deal with than unfair evaluations cast individually~\cite{IEEE2102Survey}.

Although detection of unfair evaluations, mainly collusion, has been widely studied in the recent literature~\cite{mainWWW,CIKMInds,dishonest,reliable,IEEE2102Survey}, some serious challenges remain. Generally, existing collusion detection techniques rely on either using temporal and behavioral indicators to identify colluding groups~\cite{mainWWW,CIKMInds} or on continuous monitoring of the behavior of the system, looking for unusual behavior~\cite{dishonest,reptrap}.

Identifying colluding groups usually involves looking for cliques;  however, relying on solving or approximating such NP-hard problems clearly degrades the performance of these systems. Heuristic approaches, such as the Frequent Itemset Mining technique~\cite{FIM} used in~\cite{mainWWW}, do not solve the problem either, without impacting the accuracy of the system. Tuning indicators or monitoring systems in order to make them sufficiently sensitive to collusion attacks, without an excessive number of false alarms, can be a daunting task, usually relying on machine learning techniques. However, preparing an adequate training dataset for such systems is yet another serious challenge \cite{mainWWW}.

Moreover, existing collusion detection systems generally use local quality metrics, i.e., metrics which are calculated from a small subset of data, and are thus vulnerable to manipulation. The more global the metric is, the more effort is needed to manipulate it. In other words, in order to manipulate a quality metric which is calculated based on the behavior of the entire community, one must change the sentiment of the community; for online communities with millions of users this might be an intractable task. For example, the dependence of the PageRank of any particular web page on the PageRanks of virtually all other webpages on the Internet is what makes the PageRank robust against manipulation \cite{pagerank}.

In this paper we propose a method called \emph{`Rating-through-Voting'} to help address the above problems. Our method first reduces the problem of rating of a product to an ``election''. In such an election, users' feedback is seen as a vote on the most appropriate rating level for the product from a list of a few (usually at most ten) available levels. We propose an iterative voting algorithm which for each level in each election list calculates how credible it is to be accepted as the community sentiment about the product.  Using such calculated credibility degrees, we assign each voter a trustworthiness degree reflecting to what extent she has behaved in accordance with the community sentiment. In the next round of iteration we employ these trustworthiness ranks to recalculate such credibility degrees for each level. We iterate this process until it converges, i.e., until new calculated credibility degrees are sufficiently close to the previous ones. The convergence of the algorithm is guaranteed with the existence of a fixed point of a continuous mapping, which also happens to be a stationary point of a constrained optimization objective function.

Such credibility degrees of rating levels are then aggregated into rating scores of products in a fully decoupled way, allowing the system to choose aggregation method best suited to the intended application. We have tested our method both on simulated data involving very large collusion attacks as well as on real world movie rating data.

The remainder of the paper is organized as follows. A general overview of our method along with some basic definitions is in Section~\ref{sec:prel}. In Section~\ref{sec:vote} we describe the details of our voting algorithm and its use for rating  (`Rating-through-Voting'). In Section~\ref{sec:impl} we present an application scenario with its implementation details; in Section~\ref{eval} we discuss the evaluation results of our method. Section~\ref{sec:rel} is devoted to some of the related work; our conclusions are presented in Section~\ref{sec:con}.

\section{Preliminaries}\label{sec:prel}

\subsection{Definitions}
Since some of our terminology, such as ``a voter'' or ``an election'', is also used either in everyday English or other technical fields, we now specify the intended meaning of such terms as we use them in this paper.

\vspace{.3cm}
\noindent An \textbf{evaluation} is a feedback given by a person on a product. We call the person who casts the evaluation a \emph{rater} or a \emph{voter}. In order to avoid overusing a term, we may use words evaluation, vote and feedback synonymously.

\vspace{.3cm}
\noindent An \textbf{election} is the process of choosing one item from a finite list of items. When we speak about rating through voting, the election items on a voting list are the possible rating levels for the quality of a product; for example, $1$-$10$ for rating movies in IMDb or $1-5$ to represent quality of a product in Amazon online market.

\vspace{.3cm}
\noindent \textbf{Trustworthiness} of a voter is a metric which shows to what extent the voter has been voting in accordance with the sentiment of the community.

\vspace{.3cm}
\noindent \textbf{Credibility} of a an item on an election list is a metric which shows the level of community approval of that item.

\vspace{.3cm}
\noindent \textbf{Rating Score} are the scores which are produced by our system or other systems which we use for comparison purpose to reflect the quality of a product. We may use words `rating scores', `scores' or `ranks' interchangeably to refer to such scores.
\vspace{.3cm}

Of course, the precise meaning of the above terms will become apparent only after we present their technical usage.

\subsection{Method Overview}

Our method is based on the idea of reducing rating to voting. In rating systems we have a list of products to be rated by a group of users. Every users can express her opinion on the quality of the product by casting an evaluation of the quality level which describes the best the opinion of the user. For example, in Figure~\ref{fig:r}, three users have evaluated a product by casting their evaluations.

In most rating systems the assigned rating score is a (possibly weighted) average of the cast votes. However, we can also look at rating process as an election. In such an election, the voters are the people who are evaluating the product; the candidates are all possible values which a voter can choose to express her opinion about the product; for example, $1$ star to indicate a low quality or $5$ stars to indicate a high quality. We will use a technical term `an item' to refer to such candidates. Our algorithm aims to determine the winner of such an election in a robust way while reflecting the prevailing opinion of the community on the quality of the product. The Figure~\ref{fig:v} depicts such an election.

Therefore, for each product in the system, we generate one election and all people who have rated that product are considered as having voted in the corresponding election. In a `classic election', the candidate who has received the majority of the cast votes will be the winner of the election. However, to make our elections robust against collusion attacks, we evaluate voters based on how their choices are supported by other community members. The closer the votes of the voter to some form of a `community consensus', the higher her trust rank is. In aggregating the votes, rather than considering them all equal, each voter's vote has the value equal to the voters trust rank. Such obtained trust ranks of voters and credibility levels of the candidates can be now used to compute the rating scores of products. We do not presuppose any particular method for computing such rating scores; such computation can be done by freely choosing a method best suited for the specific application. In our tests we have used one of many such possible options, see Section~\ref{sec:rate}.

\section{Rating-through-Voting}\label{sec:vote}
\subsection{Basic Concepts and Notation}

\noindent\textbf{Setup:} We will assume that a set of $N$ \emph{voters} $V_1,\dots,V_N$ are given a collection of  $L$ lists $\Lambda_1,\ldots,\Lambda_L$; each list $\Lambda_l$ contains $n_l$ many items $\Lambda_l=\{I_1^l,\ldots, I_{n_l}^l\}$, and the voters are asked to chose the ``best" item on each list. Not every voter is obliged to vote for the best item on every list, but can choose to vote on a subset of these lists.\\

\noindent\textbf{Problem:}  As the system receives these votes, the task is to assess:
\begin{enumerate}
\item trustworthiness of all voters;
\item level of ``community approval" for each item.
\end{enumerate}

In order to make such estimate of the level of ``community approval" robust against  possible unfair voting practices of the participants, the assessment of the trustworthiness of voters should have the following features (at the moment we allow these features to be specified both very vaguely and in a somewhat circular way).
\begin{enumerate}
\item voters who vote on a large number of lists, and whose choices are largely in agreement with the prevailing sentiment of the community of voters, should obtain higher level of trustworthiness than the voters who vote on only a few of these lists, or vote inconsistently with the prevailing sentiment;
\item voters who seldom vote or voters who vote more or less randomly, just to be seen as active in the community, and then choose to vote unfairly on a few particular lists for their choices which are not favored by the remaining voters, should not be able to secure election of their choices even if such colluding voters are a large majority for those particular lists.
\end{enumerate}

Note that it is not necessary that the voters vote simultaneously for all lists; in fact, the outcomes of voting on all but one list might have already been determined. We want to decide the outcome for the latest election using the past voting pattern of the voters (but NOT the outcomes of the past voting).

\subsection{Vote Aggregation Algorithm}

Let us first introduce some notation:
\begin{itemize}
\item
$r\rightarrow li$ denotes the fact that voter $V_r$ has participated in voting for the best object on  list $\Lambda_l$ and has chosen item $I_i$;
\item
$n_l$ denotes the number of items on list $l$;
\item for each item $I^l_{i}$ on list $\Lambda_l$ we will keep track of its \emph{level of community approval} at the step of iteration $p$, denoted by  $\rho^{(p)}_{li}$;
\item these individual community approval levels $\rho^{(p)}_{li}$ will be collected into a single vector \[\vec{\rho} = \langle \rho_{li}\; : \; 1\leq l\leq L,\; 1\leq i\leq n_l \rangle;\] thus, if we let $M= \sum_{1\leq l \leq L} n_l$, then $\vec{\rho} \in \Rset^M$;
\item we define $(\vec{\rho})_l$ to be the projection $\langle \rho_{li}\; : \; 1\leq i\leq n_l\rangle$ of $\vec{\rho}$ to the subspace corresponding to a single list $\Lambda_l$.
\item for each voter $V_r$ we will also keep track of his  \emph{trustworthiness} $T_{i}^{(p)}$ at the stage of iteration $p$.

\item for each $p\geq1$ we denote by $\|\vec{x}\|_p$ the usual $p$-norm of the vector $\vec{x}=\langle x_1,\ldots,x_n\rangle$, i.e.,
\[\|\vec{x}\|_p =\left(\sum_{i=1}^{n}x_i^p\right)^{\frac{1}{p}}.\]
\end{itemize}
\vspace{.5cm}

\begin{algorithm}[!t]
\caption{Adaptive Voting Algorithm}
\label{alg:vote}

\noindent\textbf{Initialization:} Let $\varepsilon>0$ be the precision threshold, $\alpha\geq 1$ a discrimination setting parameter and
\begin{align*}
T_{i}^{(0)}&=1;\\
\rho^{(0)}_{li}&=\frac{\sum_{r\,:\,r\rightarrow li}T_{r}^{(0)}}{\sqrt{\sum_{1\leq j\leq n_l}\left(\sum_{r\,:\,r\rightarrow lj}T_{r}^{(0)}\right)^2}}\\
&=\frac{|\{r\,:\,r\rightarrow li\}|}{\sqrt{\sum_{1\leq j\leq n_l}|\{r\,:\,r\rightarrow lj\}|^2}}.
\end{align*}
~\\

\noindent\textbf{Repeat:}\\
\begin{align}
T_{r}^{(p+1)}&=\sum_{l,i\,:\,r\rightarrow li}\rho_{li};\label{rec1}\\
\rho^{(p+1)}_{li}&=\frac{\sum_{r\,:\,r\rightarrow li}\left(T_{r}^{(p+1)}\right)^{\alpha}}{\sqrt{\sum_{1\leq j\leq n_l}\left(\sum_{r\,:\,r\rightarrow lj}\left(T_{r}^{(p+1)}\right)^{\alpha}\right)^2}};\label{rec2}
\end{align}\\
\noindent\textbf{until:} $\|\vec{\rho}^{\;(p+1)}- \vec{\rho}^{\;(p)}\|_2< \varepsilon$.\\

\end{algorithm}

Algorithm~\ref{alg:vote} shows our voting algorithm. We now explain the intuitive motivation behind it.  We start with the usual vote count, where every voter has one, equally worth vote as every other voter. For each item $I^l_i\in \Lambda_l$ on a list $\Lambda_l$, the initial rank $\rho_{li}^{(0)}$ of $I^l_i$ is simply the number of votes which this item has received, normalized so that $\sum_{j}\rho_{lj}^2=1$. In the next round of iterative aggregation of the votes, each voter first gets its trustworthiness rank, which is the sum total of the ranks of all the items which he has voted for, ``prorated'' by a monotonically increasing function $f(x)=x^{\alpha}$; we will later explain such choice of $f(x)$.

The idea is that now voters themselves can be judged by the selections they have made; a high trustworthiness rank will be given only to voters who have often chosen candidates favored by many other members of the community, thus voting in accordance with the prevailing community sentiment. Such voters can be considered as ``reliable voters", choosing candidates in accordance with the community sentiment. On the other hand, those who badly judge others will themselves receive a low trustworthiness rank. We could not help mentioning that this idea is remarkably old: ``Judge not, and you shall not be judged ...'' as well as ``Judge not, that you be not judged'' (The New Testament, Luke 6:37 and Matthew 7:1, respectively).

Now we recalculate the ranks of items using \eqref{rec2}; thus each received vote is now worth the present trustworthiness rank of the voter giving such vote. We continue such iterations until the ranks stop changing significantly, i.e., we stop when $\|\vec{\rho}^{\;(p+1)}- \vec{\rho}^{\;(p)}\|_2<\varepsilon$, where $\varepsilon$ is a threshold corresponding to a desired precision adequate for the particular application; in our experiments it was in the range $10^{-6}-10^{-12}$ with the algorithm terminating after $10-40$ iterations.

The value of the parameter $\alpha$ determines the robustness of our algorithm against unfair voters. Clearly, higher values of $\alpha$ increasingly favor voters with a high compliance with the prevailing community sentiment and penalize harsher for votes given to less favoured candidates. While this makes our system more robust, large values of \ $\alpha$\  increasingly marginalize a significant fraction of honest, but less successful voters. In our experiments the values $\alpha <1.5$ were insufficient to obtain satisfactory robustness; the values $1.5<\alpha<3$ gave excellent performance without marginalizing a large number of voters, with the value $\alpha=2$ chosen for our implementation of the \emph{MovieTrust} (see Section~\ref{sec:impl}).

Our algorithm will eventually terminate with  any strictly increasing, twice continuously differentiable function $f(x)$ in place of $x^\alpha$, but we saw little value in such choices. As it will be obvious from the convergence proof below, the choice $f(x)=x^\alpha$ has a natural motivation, defining a commonly used norm on the vector space of trustworthiness ranks. We now rigorously prove that our algorithm converges and characterize the values $\vec{\rho}$ which it produces.

\subsection{Convergence proof}
\begin{figure*}
\centering
\subfigure[case when $h$ is sufficiently small]{
\includegraphics[scale=0.5]{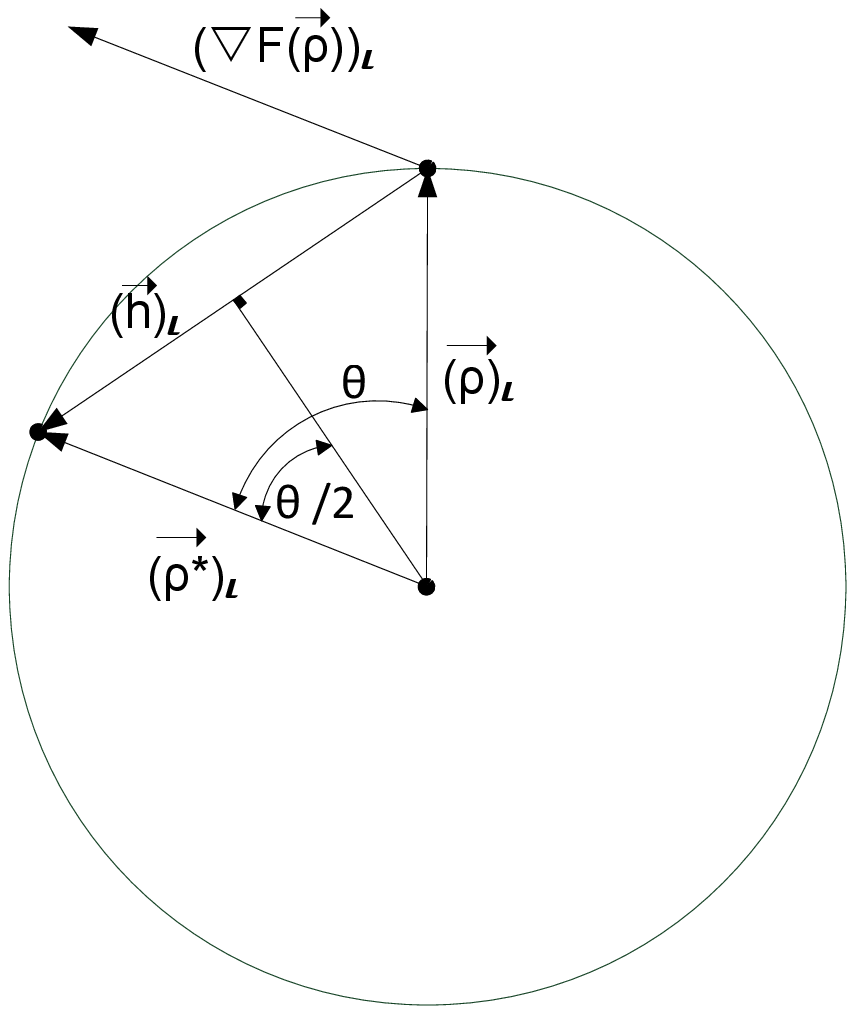}
\label{fig:c1}
}\hspace{12em}
\subfigure[additional initial line search]{
\includegraphics[scale=0.5]{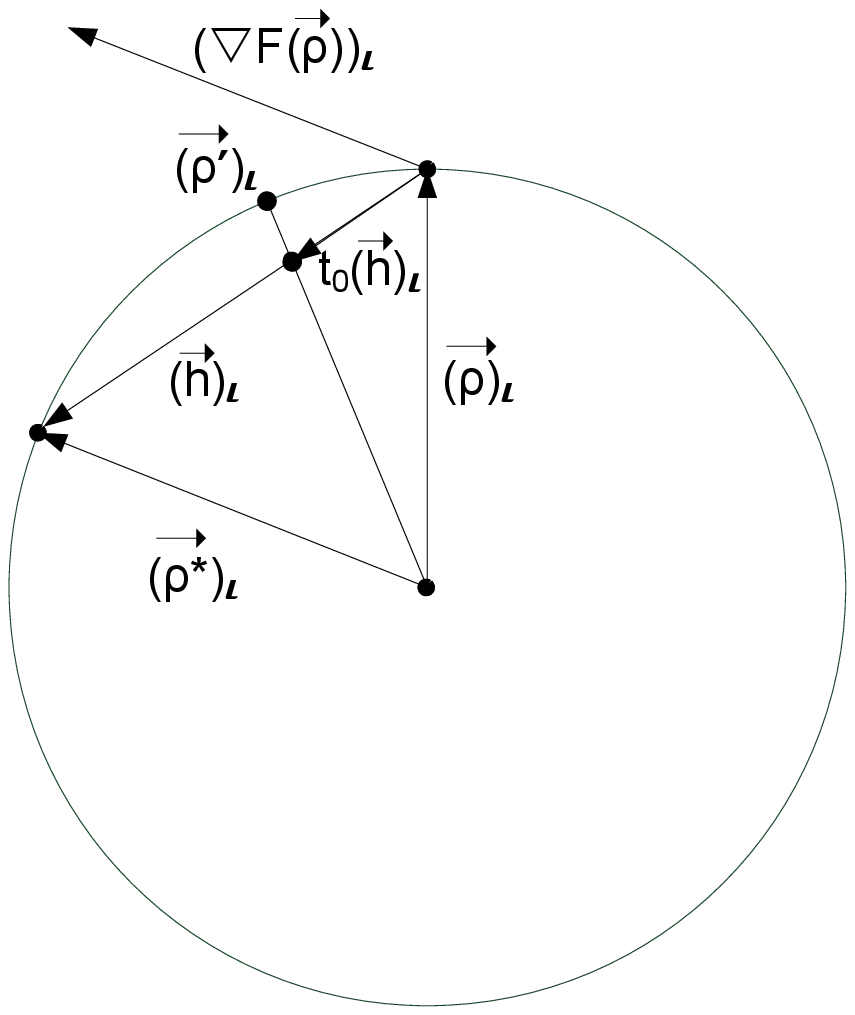}
\label{fig:c2}
}
\label{fig:c}
\caption[]{A geometric representation of the iterative procedure}
\end{figure*}
For a given `community approval' ranking $\vec{\rho}$ of items, referred in the sequel simply as  \emph{ranks of items}, let us define the corresponding \emph{trustworthiness}  $T_r(\vec{\rho})$ of a voter $V_r$ as
\begin{equation}
T_r(\vec{\rho})=\sum_{p,k\,:\,r\rightarrow pk}\rho_{pk},
\end{equation}
and denote by $\vec{T}(\vec{\rho})$ the vector of such trustworthiness ranks,
$\vec{T}(\vec{\rho})=\langle T_1(\vec{\rho}),T_2(\vec{\rho}),\ldots,T_N(\vec{\rho})\rangle$; then
\begin{eqnarray*}&\|\vec{T}\|_{\alpha+1}=\left(\sum_{r}T_r^{\alpha+1}(\vec{\rho})\right)^{\frac{1}{\alpha+1}}&\end{eqnarray*}
is the $\alpha+1$-norm of the vector $\vec{T}(\vec{\rho})\in \Rset^N$.

We now wish to assign the ranks to the items ranked so that:
\begin{enumerate}
\item for every list $\Lambda_l$, the vector of ranks of all items on that list is a unit vector, i.e., $\|(\vec{\rho})_l\|_2=1$;
\item the $\alpha+1$ norm $\|\vec{T}\|_{\alpha+1}$ of the trustworthiness vector $T(\vec{\rho})$ is maximized.
\end{enumerate}

The reason for considering $\|\vec{T}(\vec{\rho})\|_{\alpha+1}$ rather than $\|\vec{T}(\vec{\rho})\|_{\alpha}$  in the second condition will be clear below. Note that, in a sense, this gives ``the benefit of the doubt'' to all voters, giving them the largest possible joint trustworthiness rank (in terms of the $\alpha+1$ norm of the trustworthiness vector), while maintaining  for each list the same unit $2$-norm of the vector corresponding to ranks of all objects on that list.  Thus, if we now define
\begin{equation}
F(\vec{\rho})=(\|\vec{T}\|_{\alpha+1})^{\alpha+1},
\end{equation}
then our aim is equivalent to maximizing $F(\vec{\rho})$, subject to the constraints
\[\mathcal{C}=\left\{\sum_{i=1}^{n_l}\rho_{li}^2=1, \ \ 1\leq l\leq L\right\}.\]
For this purpose we introduce for each list $\Lambda_l$ a Lagrangian multiplier $\lambda_l$, define $\vec{\lambda}=\langle \lambda_l,\;:\; 1\leq l \leq L\rangle$ and look for the stationary points of the Lagrangian function
\begin{align*}
\Phi(\vec{\rho},\vec{\lambda})&=F(\vec{\rho})-\sum_{q=1}^{L}\lambda_q\left(-1+\sum_{m=1}^{n_q}\rho_{qm}^2\right).
\end{align*}

Let $l,m$ be list indices and $i,j$ item indices; then, using the fact that  the trustworthiness functions $T_r(\vec{\rho})$ are linear, we obtain
\begin{align}
\frac{\partial F(\vec{\rho})}{\partial\rho_{li}}&=
(\alpha+1)\sum_{r\,:\,r\rightarrow li}T_r^{\alpha}(\vec{\rho});\label{grad}\\
\frac{\partial \Phi(\vec{\rho},\vec{\lambda})}{\partial\rho_{li}}&=
(\alpha+1)\sum_{r\,:\,r\rightarrow li}T_r^{\alpha}(\vec{\rho})-2\lambda_l\rho_{li}\\
\frac{\partial^2 F(\vec{\rho})}{\partial\rho_{li}\partial\rho_{mj}}&=
\alpha(\alpha+1)\sum_{r\,:\,r\rightarrow li,mj}T_r^{\alpha-1}(\vec{\rho}).
\end{align}
Note that, since every voter votes for at most one item on each list,
$\frac{\partial^2 F(\vec{\rho})}{\partial\rho_{li}\partial\rho_{lj}}=0$ for every $l,i,j$ such that $i\neq j$.
Also, if $(\vec{\rho},\vec{\lambda})$ is a stationary point of $\Phi$ then $\frac{\partial \Phi(\vec{\rho},\vec{\lambda})}{\partial\rho_{li}}=0$ and thus
\begin{align}\label{r1}
{\rho_{li}}{\lambda_l}=\frac{\alpha+1}{2}\sum_{r\,:\,r\rightarrow li}T_r^{\alpha}(\vec{\rho}).
\end{align}
This yields
\begin{align*}
\rho_{li}^{2}\lambda_l^2=\frac{(\alpha+1)^2}{4}\left(\sum_{r\,:\,r\rightarrow li}T_r^{\alpha}(\vec{\rho})\right)^2,
\end{align*}
and by summing the above equations for all indices $i$ of objects on the list $l$ we get
\begin{align*}
{\lambda_l}^2\sum_{i=1}^{n_l}\rho_{li}^{2}=\frac{(\alpha+1)^2}{4}\sum_{m}\left(\sum_{r\,:\,r\rightarrow lm}T_r^{\alpha}(\vec{\rho})\right)^2.
\end{align*}
Since $(\vec{\rho},\vec{\lambda})$ is a stationary point of $\Phi$ also $\frac{\partial \Phi(\vec{\rho},\vec{\lambda})}{\partial\lambda_{l}}=0$; this implies
$\sum_{i=1}^{n_l}\rho_{li}^{2}=1$, and since by \eqref{r1} $\lambda_l$ must be positive, we obtain from the above and from \eqref{r1}
\begin{align}\label{fixed}
\rho_{li}=\frac{\sum_{r\,:\,r\rightarrow li}T_r^{\alpha}(\vec{\rho})}{\sqrt{\sum_{m}\left(\sum_{r\,:\,r\rightarrow lm}T_r^{\alpha}(\vec{\rho})\right)^2}}.
\end{align}

We now define
$\vec{\rho}\mapsto (\vec{\rho})^\ast$ to be the mapping such that for an arbitrary $\vec{\rho}$,
\begin{equation}\label{over}
(\vec{\rho})_{li}^\ast=\frac{\sum_{r\,:\,r\rightarrow li}T_r^{\alpha}(\vec{\rho})}{\sqrt{\sum_{m}\left(\sum_{r\,:\,r\rightarrow lm}T_r^{\alpha-1}(\vec{\rho})\right)^2}}.
\end{equation}
Recall that we denote by $(\vec{x})_{l}$ the projection of a vector $\vec{x}\in\Rset^M$ to the subspace of dimension $n_l$, which corresponds to a list $\Lambda_l$. Thus, using \eqref{grad}, equations \eqref{over} can be written as
\begin{equation}\label{str}
(\vec{\rho}^{\,\ast})_{l}=\frac{(\nabla F(\vec{\rho}))_l}{\| (\nabla F(\vec{\rho}))_l \|_2}.
\end{equation}
Consequently, $(\vec{\sigma},\vec{\lambda})$ is a stationary point of $\Phi$ just in case $\vec{\sigma}^\ast=\vec{\sigma}$, i.e., for all $1\leq l\leq L$,
\begin{equation}
(\vec{\sigma})_l=\frac{(\nabla F(\vec{\sigma}))_l}{\| (\nabla F(\vec{\sigma}))_l \|_2}.
\end{equation}
Note also that in our algorithm the approximation $\vec{\rho}^{\;(n+1)}$ of the vector $\vec{\rho}$ obtained at the stage of iteration $(n+1)$ can be written as
\[
\vec{\rho}^{\;(n+1)}=(\vec{\rho}^{\;(n)})^\ast,
\]
and that our algorithm will halt when $\vec{\rho}^{\;(n)}$ get sufficiently close to a fixed point $\vec{\sigma}=(\vec{\sigma})^{\;\ast}$ of the mapping $\vec{\rho}\rightarrow (\vec{\rho})^{\;\ast}$, a stationary point of the Lagrangian function $\Phi(\vec{\rho},\vec{\lambda})$. As we will see, such a point $\vec{\sigma}$ is a constrained local maximum of $F(\vec{\rho})$, subject to the constraints $\|(\vec{\rho})_l\|=1$, \ $1\leq l\leq L$. Note that we do not need to prove the uniqueness of such a fixed point because our final ranks are \emph{defined} as the outputs of our algorithm, and we only need to prove that our algorithm eventually terminates; for this purpose  just approaching \emph{a fixed point} is sufficient.

Let $\vec{\rho}$ be an arbitrary vector such that  $\|(\vec{\rho})_l\|_2=1$ for all $1\leq l\leq L$; we abbreviate  $(\vec{\rho})^{\ast}$ with $\vec{\rho}^{\;\ast}$ and let $\vec{h}=\vec{\rho}^{\;\ast}-\vec{\rho}$. By applying the Taylor formula with the remainder in the Lagrange form,
we get that for some $0< c <1$ and $\vec{\mu}_c= c \vec{\rho}+(1-c)\vec{\rho}^{\,\ast}$
We have
\begin{align}\label{taylor}
F(\vec{\rho}^{\,\ast})&=F(\vec{\rho}+\vec{h})\nonumber\\
&=F(\vec{\rho})+
\nabla F(\vec{\rho})\cdot \vec{h}+\frac{1}{2}\sum_{l,m,i,j}\frac{\partial^2 F(\vec{\mu}_c)}{\partial\rho_{li}\partial\rho_{mj}}h_{li}h_{mj}.
\end{align}
Since
\begin{align}
(\vec{h})_l&=(\vec{\rho}^{\,\ast})_l-(\vec{\rho})_l=\frac{(\nabla F(\vec{\rho}))_l}{\| (\nabla F(\vec{\rho}))_l \|_2}-(\vec{\rho})_l,
\end{align}
using also \eqref{str}, we get
\begin{align*}
(\nabla F(\vec{\rho}))_l\cdot (\vec{h})_l&=(\nabla F(\vec{\rho}))_l\cdot\left(\frac{(\nabla F(\vec{\rho}))_l}{\| (\nabla F(\vec{\rho}))_l \|_2}-(\vec{\rho})_l\right)\\
&=\| (\nabla F(\vec{\rho}))_l \|_2-(\nabla F(\vec{\rho}))_l\cdot(\vec{\rho})_l\\
&=\| (\nabla F(\vec{\rho}))_l \|_2-\|(\nabla F(\vec{\rho}))_l\|_2(\vec{\rho}^{\,\ast})_l\cdot(\vec{\rho})_l\\
&=\| (\nabla F(\vec{\rho}))_l \|_2(1-(\vec{\rho}^{\,\ast})_l\cdot(\vec{\rho})_l)
\end{align*}
Let $\theta_l$ be the angle between the unit vectors $(\vec{\rho})_l$ and $(\vec{\rho}^{\;\ast})_l$, i.e., such that $\cos \theta_l = (\vec{\rho})_l\cdot(\vec{\rho^*})_l$. Then, (see Figure~\ref{fig:c1})
\[
\left\|\frac{(\vec{h})_l}{2}\right\|^2_2=\left(\sin\frac{\theta_l}{2}\right)^2=\frac{1-\cos\theta_l}{2}=
\frac{1-(\vec{\rho})_l\cdot(\vec{\rho}^\ast)_l}{2}.
\]
Combining the last two formulas we get
\begin{align}\label{ll}
(\nabla F(\vec{\rho}))_l\cdot (\vec{h})_l&=
\frac{\| (\nabla F(\vec{\rho}))_l \|_2\;\|(\vec{h})_l\|^2_2}{2}.
\end{align}
Assume first that $\|\vec{h}\|_2$ is sufficiently small, so that the contribution of the second order terms in \eqref{taylor} is small compared to the first order term, and, consequently
\begin{align}\label{l4}
F(\vec{\rho}^{\,\ast})\approx F(\vec{\rho})+\nabla F(\vec{\rho})\cdot \vec{h}.
\end{align}
Since $\|\nabla F(\vec{\rho})\|_2$ is a continuous function, it achieves its minimum on the compact set defined by our constraints, i.e., on the set
$\mathcal{C}=\{ \vec{\rho}\; : \; \|(\vec{\rho})_l\|=1, \ 1\leq l \leq L\}$. It is easy to see that the directional derivative of  $F(\vec{\rho})$ in the (radial) direction of vector $\vec{\rho}$ is always strictly positive; thus, on the compact set defined by our constraints its minimum must also be strictly positive. Thus, there exists $\kappa >0$ such that $\|\nabla F(\vec{\rho})\|_2>\kappa$ for all $\vec{\rho}\in \mathcal{C}$; using this and by summing equations \eqref{ll} for all $1\leq l\leq L$, we get
\begin{align}\label{lll}
\nabla F(\vec{\rho})\cdot \vec{h}&>
\frac{\kappa\;\|\vec{h}\|^2_2}{2}.
\end{align}
This and  \eqref{l4} imply that, for $\vec{\rho}^{\;(n)}$ and $\vec{h}^{\;(n)}=\vec{\rho}^{\;(n+1)}-\vec{\rho}^{\;(n)}$ obtained in our iterations,
\[F(\vec{\rho}^{\;(n+1)})-F(\vec{\rho}^{\;(n)})=F((\vec{\rho}^{\;(n)})^\ast)-F(\vec{\rho}^{\;(n)})> \frac{\kappa\;\|\vec{h}^{\;(n)}\|^2_2}{2}\]
Consequently, since $F(\vec{\rho})$ must be bounded on a compact set defined by the constraints $\mathcal{C}$, we get that  $\|\vec{h}^{\;(n)}\|^2_2$ must converge to zero, i.e.,  $\|\vec{\rho}^{\;(n)}-(\vec{\rho}^{\;(n)})^\ast\|_2$ will eventually be smaller than the prescribed threshold and the algorithm will terminate.

If $\|\vec{h}\|_2$ is not sufficiently small so that the impact of the second order term in \eqref{taylor} makes the inequality \eqref{lll} false, we supplement our algorithm with an initial phase which involves a line search. While this ensures a provable convergence of our algorithm, in all of our (very numerous) experiments such a line search was never activated; however, we were unable to prove without any additional assumptions that indeed such line search is superfluous, so we present a slight modification of our algorithm. Let
\[
f(\vec{\rho},t)=F(\vec{\rho}+t(\vec{\rho}^{\; \ast}-\vec{\rho}));
\]
then, by the previous considerations, for sufficiently small $t$ function $f(\vec{\rho}^{\;(n)},t)$ is increasing in $t$. We now modify our iteration step as follows.
If there exists $t_0\in (0,1)$ such that $\frac{\partial f}{\partial t}(\vec{\rho},t_0)=0$ (testing this amounts to solving a low degree algebraic equation), then we let $\vec{\rho}^{\;(n+1)}=\vec{\rho}^{\;\prime}$, where $\vec{\rho}^{\;\prime}$ is defined so that for all $1\leq l\leq L$, and for the smallest root $t_0$ of the above equation,
\begin{align*}
(\vec{\rho}^{\;\prime})_l&=\frac{(\vec{\rho}^{\;(n)})_l+t_0((\vec{\rho}^{\;(n)\; \ast})_l-(\vec{\rho}^{\;(n)})_l)}{\|(\vec{\rho}^{\;(n)})_l+t_0((\vec{\rho}^{\;(n)\; \ast})_l-(\vec{\rho}^{\;(n)})_l)\|_2};\hspace*{5mm}
\end{align*}
see Figure~\ref{fig:c2}; if no such $t_0$ exists, we let
\begin{align*}\vec{\rho}^{\;(n+1)}&=(\vec{\rho}^{\;(n)})^{ \ast}.
\end{align*}
The convergence now follows from an argument similar to the one in the previous case.

\subsection{Rating Through Voting Procedure}\label{sec:rate}
We now describe how we use the proposed voting algorithm in online rating systems.

Our rating procedure starts by assigning to each product $\pi_l$, $1\leq l\leq L$, a voting list $\Lambda_l$; the items on each voting list are the rating levels, comprising  a ``scale'' from $1$ to $n$, (in practice $n\leq 10$). Each rater is now construed as a voter $V_r$, ($1\leq r\leq N$).  We then process cast evaluations, by interpreting an evaluation of level $i$,  ($1\leq i\leq n$), of a product $\pi_l$, ($1\leq l\leq L$),  by a voter $V_r$,  as his vote for item $i$ on the list $\Lambda_l$. After our iterative algorithm has terminated, each level $1\leq i\leq n$ on each voting list $\Lambda_l$ has received a corresponding credibility degree $\rho_{li}$.

We can now obtain a rating score $\mathrm{R}(\pi_l)$ of each product $\pi_l$, using such credibility degrees $(\vec{\rho})_l=\langle \rho_1,\ldots,\rho_n\rangle$ in a way which suits the particular application best. For example, if the rating scores have to reflect where the community sentiment is centered, we can simply choose as the rating score $\mathrm{R}(\pi_l)$ of $\pi_l$ the rating level which has the highest credibility rank. Such a rating score does not involve any averaging and is most indicative of the community's \emph{prevailing sentiment}. On the other hand, if we wish to obtain a score which emphasises such prevailing sentiment, but, to a varying degree takes into account ``dissenting views'', one can form a weighted average of the form
\[\mathrm{R}(\pi_l)=\sum_{1\leq i\leq n_l}\frac{\rho_{li}^p\;\times i}{\sum_{1\leq j\leq n_l}\rho_{lj}^p}.\]
where $p\geq 1$ is a parameter. As $p$ increases, such rank converges to the previous, ``maximum credibility'' rank, while for smaller values of $p$ we obtain a significant averaging effect.  In our implementations and testing of the MovieTrust we have used  $p=2$.

\section{Implementation}\label{sec:impl}

\subsection{Application Scenario}
\label{sec:ascen}
One of the most popular rating scenarios on the web is movie rating. There are several movie rating systems which allow their users to rate movies and post their reviews. Based on users' evaluations of the movies, a system can calculate a rating score for every movie. Such rating scores are usually some forms of average of the evaluations posted by users. For example, IMDb is one of the best known online movie rating systems. It uses an algorithm for calculating rating scores for movies which the web site owners do not disclose, wishing, as they declare,  to keep it effective, but they explicitly say that their ranks are a weighted average~\cite{IMDBR}.

Another movie rating and recommending system is MovieLens~\cite{mlens} provided by  \emph{GroupLens} research lab at the University of Minnesota~\cite{glens}. MovieLens uses a collaborative filtering method to recommend movies to its users, based on their personal preferences. In this paper we test our system using a partial copy of MovieLens data obtained form GroupLens website~\cite{dataset}. The dataset contains $855598$ rating scores cast by $2113$ users on nearly $10000$ movies. This dataset for every movie also contains the corresponding rating score given by the top critics of \emph{Rotten Tomatoes} movie rating website~\cite{rt}, another well known movie rating system which we also use for comparison purposes.

\subsection{MovieTrust}

To test our methodology, we have designed and implemented a movie rating system, which we call \emph{MovieTrust}, aiming to robustly calculate rating scores for movies. MovieTrust comprises of three components. One is the rating calculation engine, the backbone of the application,  which calculates the rating scores based on locally stored data. The second component is an API which makes MovieTrust services available to users all around the Internet. The third component is an extension for the Google Chrome browser. At the moment, this extension is developed specifically for IMDb website.

\begin{figure}
\centering
\includegraphics[scale=0.22]{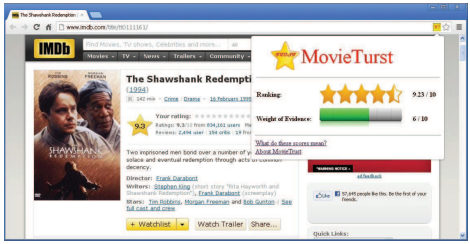}
\label{fig:shot}
\caption[]{Screen shot showing application of the MovieTrust Chrome Extension}
\end{figure}

When the MovieTrust extension is installed on the Chrome browser and the users visit IMDb  page of a movie, the MovieTrust icon will appear on the right corner of the address bar and users can simply click on it and see what rating score has been calculated for that movie by our system. Figure~\ref{fig:shot} shows a screen shot of MovieTrust extension on the Chrome browser.

The MovieTrust has a website~\cite{mtrust} which provides all related information about the tool. The extension is available for download, to allow easy evaluation of the performance of our system and comparison with the well known movie ratings system, IMDb.

\section{Evaluation}\label{eval}
We evaluate two aspects of the performance of our method: robustness against unfair evaluations and the accuracy of calculated rating scores.

\subsection{Evaluating Robustness}
We use three scenarios to evaluate the robustness of our method.
\subsubsection{Evaluation Scenarios}
\noindent \textbf{\underline{Scenario 1.}}
We start evaluation of our system with a very simple example, which nevertheless illustrates the heuristics which was the starting point for our algorithm. Table~\ref{tbl:simple} shows the votes cast by $5$ voters~($r_1,\ldots, r_5$) on $5$ items~($I_1,\ldots, I_5$) in $6$ different elections~($\Lambda_1,\ldots,\Lambda_6$). For example, the first row shows that in the first election $r_1$, $r_2$ and $r_3$ have voted for $I_1$ and $r_4$ and $r_5$  have voted for $I_2$.

In the usual election model, in the last election  $I_1$ must win because it has received $3$ votes out of $5$. However, when we look at the history of the voters' choices, $r_2$ and $r_3$, who have voted for $I_2$ in last election, have voted for the winners in all past elections. This means that they have been always behaving close to community consensus. On the other hand $r_1$, $r_4$ and $r_5$ who have voted for $I_1$ do not conform to community consensus in most of the past elections. Therefore, we can argue that in the last election the $I_2$ should win, despite of the fact that the majority have voted for $I_1$.

\begin{table*}[!t]
\centering
\caption{A simple example for showing impact of voter trustworthiness on election result(Scenario $1$).}
\vspace{0.5cm}
\label{fnru435n}
\subtable[dataset.]{
        \begin{tabular}{|l| c c c c c |}
        \hline
        \backslashbox{\footnotesize{Election}}{\footnotesize{Voter}}& $r_1$ &  $r_1$  & $r_1$ & $r_1$ & $r_1$ \\
        \hline
        $\Lambda_1$ &1& 1& 1& 2& 2  \\
        $\Lambda_2$  & 1& 2& 2& 3& 2  \\
        $\Lambda_3$  & 3& 4& 4& 4& 2  \\
        $\Lambda_4$  & 1& 3& 3& 3& 1  \\
        $\Lambda_5$  & 2& 2& 2& 1& 1  \\
        $\Lambda_6$  & 1& 2& 2& 1& 1  \\
        \hline
        \end{tabular}
    \label{tbl:simple}
}
\hspace{5em}
\subtable[Results]{
        \begin{tabular}{|l| c c c c c c |}
        \hline
        \backslashbox{\footnotesize{Item}}{\footnotesize{Election}}& $\Lambda_1$ &  $\Lambda_2$   & $\Lambda_3$  & $\Lambda_4$  & $\Lambda_5$  &$\Lambda_6$  \\
        \hline
        $I_1$ &0.99& 0.08& 0& 0.11& 0.11& 0.20\\
        $I_2$ & 0.11& 0.99& 0.03& 0& 0.99& 0.98 \\
        $I_3$ & 0& 0.09& 0.08& 0.99& 0& 0  \\
        $I_4$ & 0   & 0   & 1 & 0   & 0   & 0 \\
        $I_5$ & 0   & 0   & 0   & 0   & 0   & 0  \\
        \hline
        \end{tabular}
       \label{tbl:simpleres}
}
\end{table*}
\begin{table*}
    \begin{center}
    \caption{The Results of running our model in the presence of a large number of colluders (Scenario $2$).}
    \vspace{0.5cm}
        \begin{tabular}{|l| c c c c c c c | c|}
        \hline
        \backslashbox{Item}{Election}& $\Lambda_1$ &  $\Lambda_2$   & $\Lambda_3$  & $\Lambda_4$  & $\Lambda_5$  &$\Lambda_6$&$\Lambda_7$ & (INITIAL)  \\
        \hline
        $I_1$ & 0.9561    & 0.1662    & 0.0226   & 0.2787   & 0.2812    & 0.4542    &  \textbf{\underline{0.9586}}&0.3162 \\
        $I_2$ & 0.2714    & 0.9646    & 0.1290   & 0.0334   & 0.9569    & 0.8840    & 0                             &             \\
        $I_3$ & 0.0130    & 0.1795    & 0.1742   & 0.9556   & 0.0469    & 0.0315    & 0                             &              \\
        $I_4$ & 0.0503    & 0.0089    & 0.9731   & 0.0270   & 0.0248    & 0.0402    & 0&             \\
        $I_5$ & 0.0028    & 0.0572    & 0.0164   & 0.0606   & 0.0175    & 0.0624    &\textbf{\underline{0.2846}}  & 0.9486\\
        $I_6$ & 0.0490    & 0.0453    & 0.0235   & 0.0205   & 0.0351    & 0.0603    & 0                             &             \\
        $I_7$ & 0.0812    & 0.0638    & 0.0207   & 0.0180   & 0.0272    & 0.0386    & 0                             &           \\
        $I_8$ & 0.0164    & 0.0078    & 0.0649   & 0.0519   & 0.0069    & 0.0200    & 0                             &            \\
        \hline
        \end{tabular}
    \end{center}
  \label{tbl:secondres}
\end{table*}

Table~\ref{tbl:simpleres} shows the results of running our model on data provided in Table ~\ref{tbl:simple}. During initialisation of our algorithm, item  $I_1$ receives initial rank of $\frac{3}{\sqrt{3^2+2^2}}\approx~0.83$ while $I_2$ receives initial rank of $\frac{2}{\sqrt{3^2+2^2}}\approx~0.55$. However, the final ranks obtained after 16 iterations, shown in table \ref{tbl:simpleres}, are $0.2$ and $0.98$ respectively; thus, $I_2$ wins the last election despite of the majority voting for $I_1$.

\vspace{0.7cm}
\noindent \textbf{\underline{Scenario 2.}}
In the second scenario, we use synthetic data to show how robust our system is against unfair evaluations such as collusion. Our new dataset contains $60$ voters voting in 7 elections, with each election list having $8$ items~($I_1,\ldots, I_8$). First $15$ voters are ``honest voters" who do not cast unfair evaluations. We generate these users and their corresponding votes by replicating voters~$r_1$ to $r_5$ three times to generate $15$ honest voters~($r_1,\ldots, r_{15}$). We also generate $45$ voters who make random choices in all elections, except the last one. In the last, $7^{th}$  election, the unfair voters collude in order to manipulate the outcome of the election, trying to secure election of item $I_5$. On the other hand, all~$15$ honest voters vote for~$I_1$.

Thus, we have $6$ elections in each of which $\frac{1}{4}$ of all voters cast their evaluations according the pattern in scenario 1, while $\frac{3}{4}$ of all voters cast random votes, and one election in which the collusion attack happens. Note that the number of ``unfair" voters is $3$ times the number of fair voters.
In a normal election scenario, it is obvious that in the last election item~$I_5$ would win, having received~$3$ times the number of votes of its opponent $I_1$. In our method, during the initialisation process, item $I_1$, having obtained $15$ votes, gets the initial ranking score $\frac{15}{\sqrt{15^2+45^2}}\approx 0.32$ and item $I_5$, having obtained $45$ votes gets an initial score of $\frac{45}{\sqrt{15^2+45^2}}\approx .95$. However, after we run our algorithm (which terminates after 16 iterations) the scores are essentially reversed and $I_1$ wins the election, having obtained credibility degree $\approx 0.96$ versus the credibility degree $\approx 0.28$ obtained by $I_2$. Such an outcome is to be expected from a robust voting system. Figure~\ref{tbl:secondres} shows the ranks produced by our algorithm, and, for comparison, in the last column, the initial ranks based on (normalized) simple vote counting.

\begin{figure*}
\centering
\subfigure[RMS for Promoted Movies]{
    \includegraphics[scale=0.4]{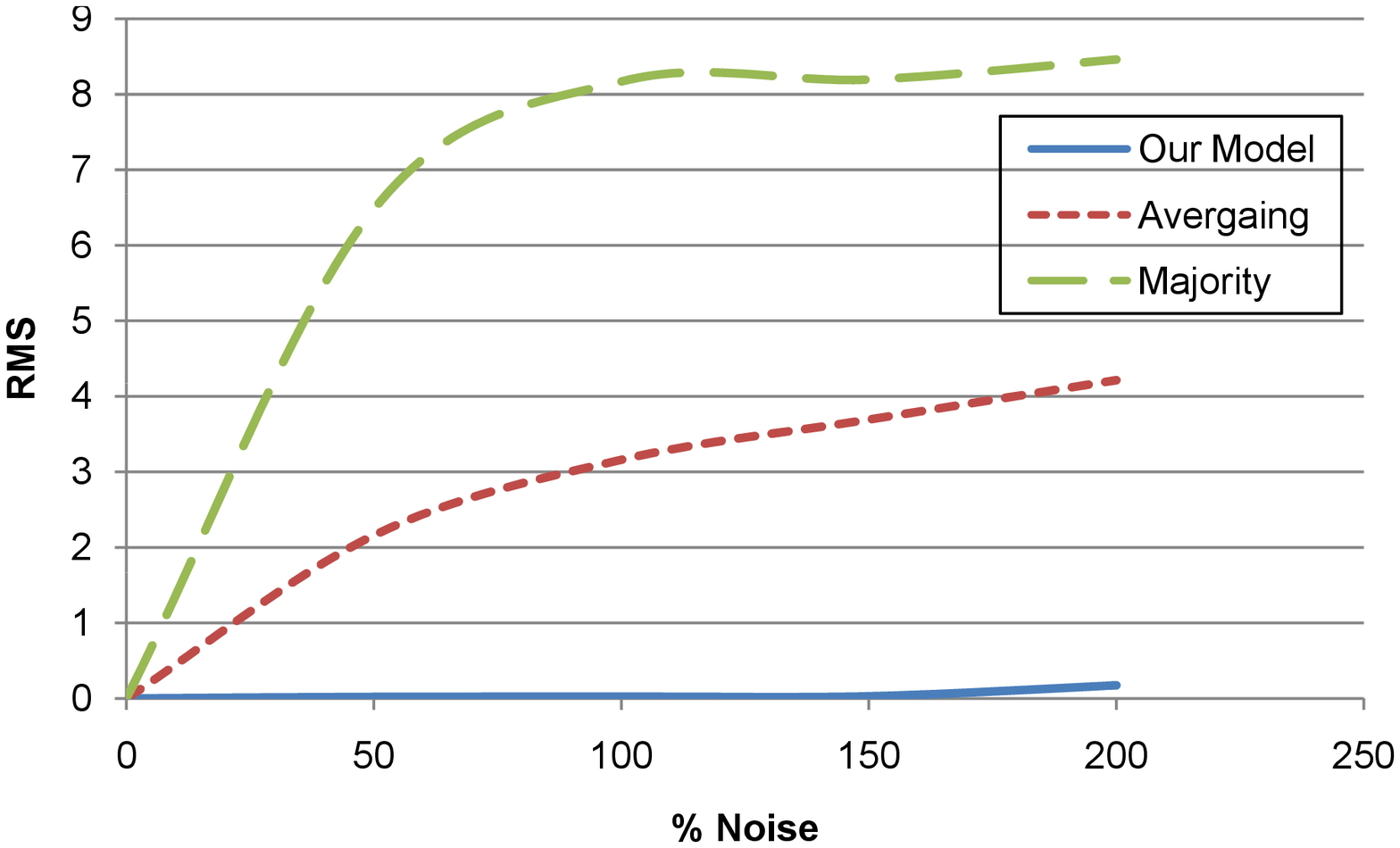}
    \label{fig:prom}
}\hspace{4em}
\subfigure[RMS for Demoted Movies]{
    \includegraphics[scale=0.4]{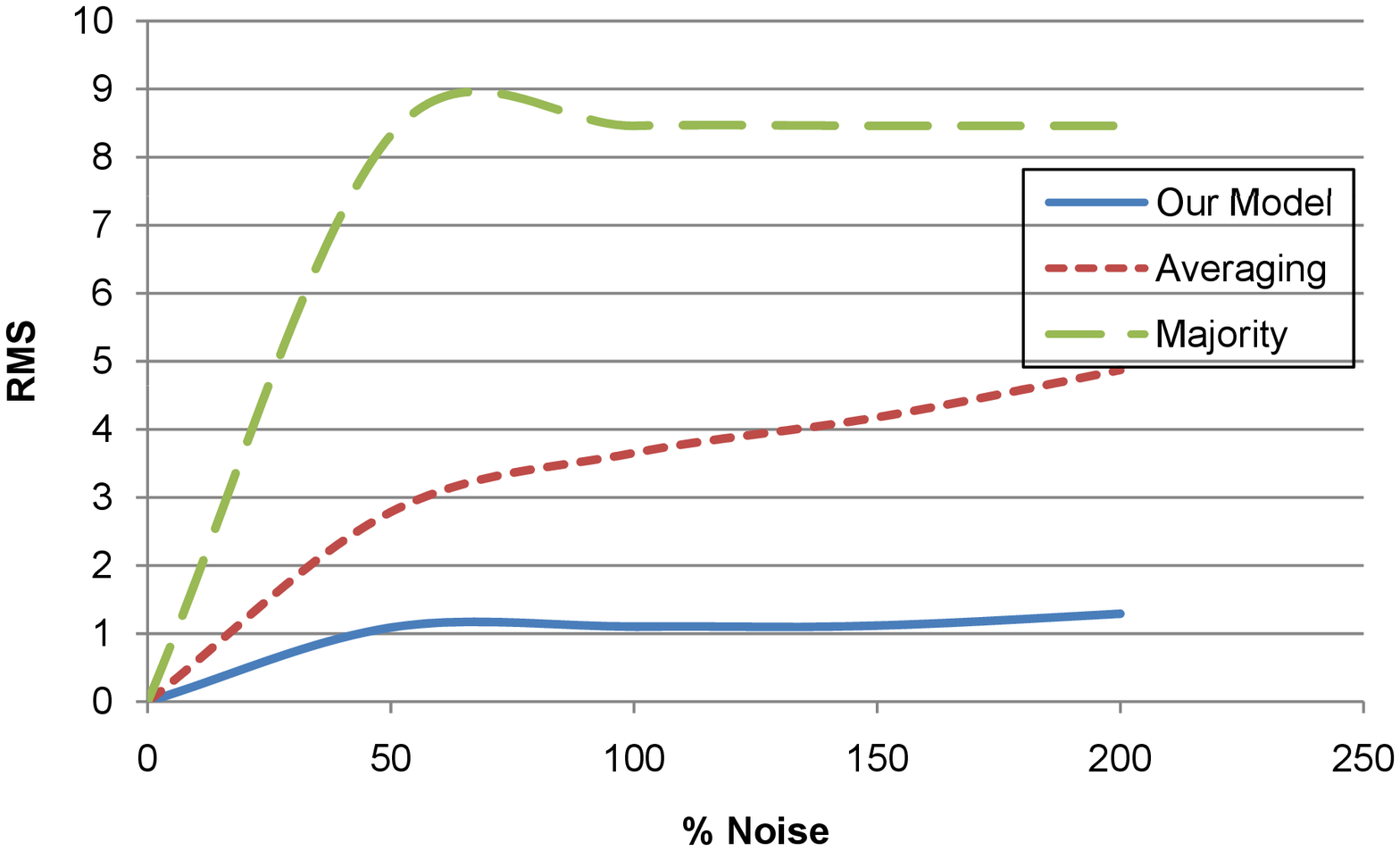}
    \label{fig:dem}
}
\caption[]{Comparison of behavior of three methods in the presence of different levels of injected collusion attacks (Scenario $3$)}
\end{figure*}

\vspace{0.7cm}
\noindent \textbf{\underline{Scenario 3.}}
In this scenario, we use real world evaluation results, from the MovieLens dataset. We inject different levels of collusion attacks to the dataset and check robustness of our method. We also compare the performance of our method with two commonly used rating models: Averaging and Majority. We are aware that there are many recent methods which try to calculate rating scores after detecting and eliminating unfair evaluations such as~\cite{mainWWW} or~\cite{dishonest}. However, it should be noted that in the presence of the massive collusion attacks which we inject to dataset, the control of the products would inevitably be given to the colluders. All existing collusion detection systems use majority as a primary indicator, such as~\cite{maj2}, or as a secondary indicator to tune their primary indicators such as~\cite{mainWWW} or~\cite{dishonest}. Therefore, they must fail when the number of unfair votes is far larger than the number of the fair ones. This explains our choice of methods used for comparison with our method.

To evaluate methods, we calculate a rating score for each movie from the data set used by applying ours, Averaging and Majority methods. We then add collusion attacks to the dataset and recalculate rating scores of movies and find the difference between new ranks and the old ranks. To quantify this differences for each method in each run, we calculate the `Root-Mean-Square (RMS)' of the differences between new ranks and old ranks for all movies.

To demonstrate the robustness of our system, we try to promote all movies which have very low rating scores, i.e., movies which majority of users have given a rating score lower than $3$, by posting evaluations with values of $10$ to sharply increase their rating scores. We also try to demote all movies having a very high rating score, i.e., movies with a rating score higher than $8$ from the majority of raters, by injecting evaluations with value of $1$ to sharply decrease the rating scores of such movies. For each movie to be promoted or demoted we inject different levels of collusion attacks, ranging from $0\%$ to $200\%$. We then compare the RMS values of the differences to show how these three methods behave in the presence of such attacks.

Suppose that for a movie $\Lambda_l$, $m_l$ evaluations have been posted by real world users. To test our method with a collusion attack of size $200\%$ of the number of the real, already existing votes, we inject $2 \times m_l$ votes which try to promote or demote the product, as explained above. This clearly creates a massive collusion attack on the product (here a movie) which sharply changes the majority consensus toward unfair evaluations, thus making every existing method likely to fail.

Figure~\ref{fig:prom}, shows the results of injecting collusion attacks in order to promote low ranked movies. As shown in the chart, the Majority model is highly vulnerable against unfair votes; its corresponding evaluation level and thus the corresponding RMS error is sharply increased to the highest possible value and remains steady at that level. The averaging model performs better, but nevertheless its RMS error steadily increases as more unfair evaluations are added. The RMS of our method is extremely small, increasing slightly as the number of unfair evaluations gets close to $200\%$, never getting close to serious malfunctioning, in sharp contrast with other two methods.

Likewise, Figure~\ref{fig:dem} shows that our model behaves well also in the presence of a massive demoting collusion. While the RMS for our method is around $1$, the Majority model jumps immediately to the highest possible level and Averaging model steadily moving towards the unfair evaluations.

The reason why the RMS error of our method is larger in the presence of demoting attacks is that for high ranked movies the total number of existing votes is much larger than the number of votes for low ranking movies targeted in promoting attacks. Consequently the number of injected collusive votes in the case of demoting attacks is much larger; for some movies the number of collusive votes exceeded $5000$ which accounts for higher overall impact on the system's RMS value. However, even with such extreme collusion attacks, our system is remarkably robust.

\subsubsection{Discussion}
We tested our system by generating groups of voters with various types of voting patterns. Our experiments show that as the number of elections increases, so does the robustness of the system, as expected. The reason is that the larger the number of elections, the harder it is to manipulate our ``globally obtained" trustworthiness ranks, in which the rank of any voter depends on ranks of all other voters. Also, as the number of items on lists increases, the possibility of casting same random votes of several voters decreases and consequently the system performs more robustly. Thus, in order to skew a particular election, colluding voters need a long history of ``honest looking" activity, or a long history of massive collusion attacks gone undetected by other, more standard  methods of collusion detection which can be used in conjunction with our method, or if the attackers actually manage to change the sentiment of the community.

To summarize, the followings are the possible scenarios in which colluders might be able to manipulate result of an election:
\begin{itemize}
  \item
        Colluders build up a large team of voters. All these voters must vote over rather long period of time in accordance with community sentiment in order to earn hight trustworthiness ranks and then attempt to manipulate a particular election. This team must be large enough to be able to dominate others while trying to promote or demote a product. This is unlikely to happen because research~\cite{mainWWW,dishonest} show that collusion happens in a short period of time and only on a few number of elections. In online rating systems, with millions of members and products, it is quite hard to collect a large fraction of members and encourage them to rate most of the existing products in the system honestly for a long period of time and then ask them to rate a particular product as the collusion team intends.
  \item
        The second possibility is to build a very large group of colluders and manipulate all elections in the system to change the sentiment of the community toward intension of collusion team. In this case since the community sentiment is toward collusion party, the honest voters will be marginalized and eventually their impact will be eliminated. This might be possible for small communities with a small number of members and items. It is obvious that it is quite impossible to run this scenario in a huge online rating system like Amazon or IMDb, and that such behavior would go undetected by other, relatively simple  collusion detection methods which easily detect such long stretches of extreme behavior and which can be deployed in parallel, with minimal additional complexity and cost of the system.
\end{itemize}

\begin{figure*}[!t]
\centering
  \includegraphics[scale=0.9]{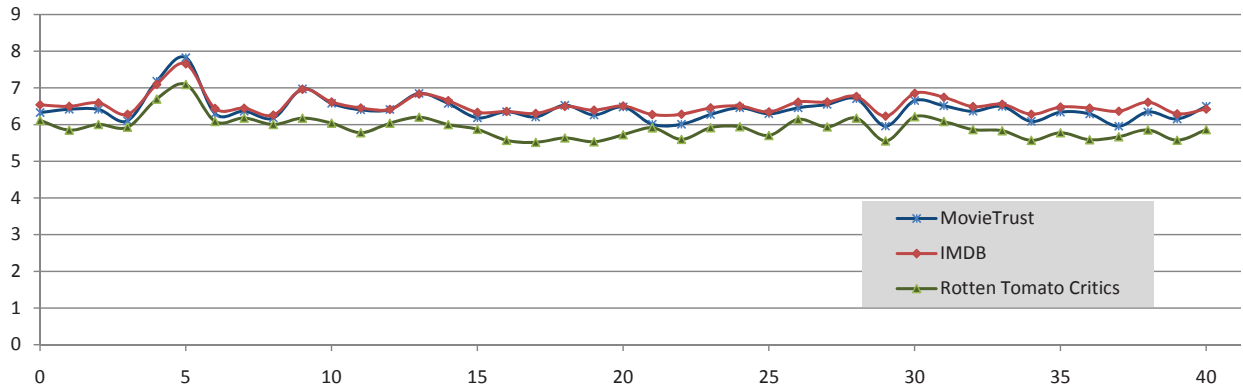}\\
  \caption{Average rating scores calculated for movies using three models: MovieTrust, IMDb and RTCritics}
  \label{fig:avg}
\end{figure*}

\subsection{Evaluating Accuracy of Results}
\subsubsection{Evaluation Scenario}
To evaluate our model we need some `gold standards' or ground truth levels to compare our calculated rating scores with and check how they deviate from such gold standards. Product rating is usually is a subjective task, i.e., the result of evaluating a product is strongly related to the personal interest and experiments of the evaluator. This subjectiveness even is getting  more serious when a movie is evaluated. It is impossible to evaluate a movie automatically and find its real quality level and compare it with our scores. On the other hand, people have different tastes and interests and the rating scores they give to a movie sometimes may be in contrast. So, it is not possible to ask a few people to watch a movie and provide us a reliable rating score for the movie as a ground truth level.

To solve this problem, we rely on rating scores calculated by two well-known existing movie rating systems. The first one is IMDb. The IMDb calculates a rating score for every movie using many parameters such as the age of the voter, the value of the posted rating score, the distribution of the votes, etc. IMDb does not disclose its method of calculating ratings, but the success of the website in the long run is a good indication that the calculated rating scores are realistic and reliable. We extracted the rating score of every movie from IMDb.

The second gold standard which we use is the rating scores given to movies by top critics in Rotten Tomatoes (referred as RTCritics in followings) movie ranking website. The critics are better informed and experienced than an average user. Although they might have their own personal interests, the community of critics as domain experts can provide dependable ratings of movies. We extracted the top critics' movie rating scores from the MovieLens dataset, as we have explained in application scenario (see Section~\ref{sec:ascen}).

First, we compare ranking scores calculated in three models. We have more than $9000$ movies in our database. We have RTCritics ranks only for about $4000$ movies but this is still too large to fit into a readable chart. So, for every $100$ movies we calculate the average score (mean). Note that this is done only to make graphical representation of rating scores of more than $4000$ moves feasible; the RMS values of the changes due to collusion attacks calculated in the previous subsection are obtained from the entire dataset, without any averaging.
We use such calculated averages to compare our rating scores with those provided by IMDb and Rotten Tomatoes. Figure~\ref{fig:avg} shows the result of this comparison. In general, our scores are higher than RTCritics and very close to the scores obtained from IMDb. The ``harsher'' ratings from Rotten Tomatoes are to be expected from a community described aptly by the label `the critics'. Moreover, not only do our ratings conform to the ratings from IMDb, but also when our ratings do differ, the difference is nearly always towards the rating scores provided by RTCritics.

\subsubsection{Discussion}

The accuracy evaluation results explicitly show that the results obtained by our system are very close to the rating scores calculated by IMDb and RTCritics. The slight differences between our method and other two are accounted for by the following reasons.
These three rankings are based on three different dataset: MovieLens (the dataset we used), IMDb dataset and Rotten Tomatoes dataset; despite the fact that the number of samples in all of these data sets is sufficiently large to expect the same statistical features, small differences still would occur, most notably because the rating scores calculated by our model are based on a dataset which holds evaluation data only up to $2009$.

More important is a high degree of similarity of the plots of the ranks produced by these three systems, see Figure~\ref{fig:avg}, strongly supporting the dictum that ``the wisdom of the crowd is reliable'', see \cite{wisdom}, in the sense that finding the community sentiment on the quality of a product is at least as dependable and, in all likelihood, more dependable than any other kind of metrics, such as various averages and the similar. This we explain by the fact that our system compensates for anomalous behavior of voters as well as for natural diversity of opinions, allowing detection of the true sentiment of the community, doing this without an external supervision or an explicit ``0-1'' classification of voters according to their perceived reliability.

In all practical systems the evaluations submitted by the participants are on a coarse granularity involving at most ten levels or less. Given the large number of voters, and relatively small number of `bins' in which their votes for each product can be placed, insures that real sentiment of the community will be captured by the emerging set of voters with high trustworthiness ranks. This insures high efficiency of our method, especially for very large online rating systems, for which trust management is both of the highest importance and one of the most challenging problems.

\section{Related Work}\label{sec:rel}
As demonstrated in~\cite{turf}, as the rating systems get more and more popular and more users rely on them to decide on purchases from  online stores, the temptation to obtain fake rating scores for products or fake reputation scores for people has dramatically increased. To detect such reviews, Mukherjee et.al.,~\cite{mainWWW} propose a model for spotting fake review groups in online rating systems. The model analyzes textual feedbacks cast on products in Amazon online market to find collusion groups. They employ FIM~\cite{FIM} algorithm to identify candidate collusion groups and then use $8$ indicators to identify colluders. In~\cite{CIKMInds} authors assign every review a degree of spam value, and based on these values they identify most suspicious users and investigate their behavior to find most likely colluders. In~\cite{unusuall} authors try to identify fake reviews by looking for unusual patterns in posted reviews.

In a more general setup, collusion detection has been studied in P2P and reputation management systems; good surveys can be found in~\cite{CollusionInP2P} and~\cite{IEEE2102Survey}. EigenTrust~\cite{eigentrust} is a well known algorithm proposed to produce collusion  free reputation scores; however, authors in~\cite{EigenCollusion} demonstrate that it is not robust against collusion. Another series of works~\cite{signaldetection,reptrap,dishonest} use a set of signals and alarms to point to a suspicious behavior. The most famous of all, the PageRank algorithm \cite{pagerank} was also devised to prevent collusive groups from building fake ranks for pages on the web.

Iterative methods for trust evaluation and ranking have been pioneered in \cite{iterativefilter1,iterativefilter2}. Some of the ideas from these papers were among the staring points of \cite{de1,de2,de3}, as the authors mention; the proof techniques which we used in this paper were inspired by the techniques developed in \cite{de1}. However, our present method sharply differs from all of these prior iterative methods by virtue of entirely  decoupling the credibility assessment from the score aggregation. More precisely, the main idea used in \cite{iterativefilter1,iterativefilter2} and in  \cite{de1,de2,de3} is  to produce at each stage of iteration an approximation of the final ratings of the objects and then calculate for each rater the degree of her ``belief divergence'' from such calculated approximations, i.e., calculate some distance measure between her proposed ranks and these approximations of the final ranks of objects. In the subsequent round of iteration a new approximation of the ranks of all objects is obtained as a weighted average of the ranks proposed by raters, with the weight given to each rater's rank inversely related to her corresponding distance from the approximate final ranks obtained in the previous round of iteration. Thus, in this manner, the ranks of objects are produced simultaneously with an assessment of the trustworthiness of the raters as reflected in the weights given to their proposed ranks.

In contrast, our iterative method operates only on credibility assessment of raters and on the levels of the community approval of items, which are obtained without using the fact that the items voted on are rating levels. In fact, as it is obvious from our algorithm, we have never used any comparisons of the proposed rating levels or even any ordering of the rating levels. We only rely on the levels of concurrence of the opinions of raters. Thus, our system can subsequently choose how to use such estimates of the `community sentiment' to produce the aggregate rating scores of items.

The second author and his co-authors, unaware of the pioneering work in~\cite{iterativefilter1,iterativefilter2,de1,de2,de3},  have proposed in \cite{aleksrep} a fixed-point algorithm for trust evaluation in online communities and subsequently in \cite{aleksmarking} an algorithm for aggregating assignment marks given by multiple assessors. This method was later applied to aggregation of sensor readings in wireless sensor networks, in the presence of sensor faults \cite{sensor}.  He also proposed the idea of applying an iterative procedure for vote aggregation to his collaborators in \cite{lazytom}; however no proof of convergence of the method was provided there, and, more importantly, the proposed method had some serious shortcomings. In the notation of the present paper, denoting again the total number of voters by $N$ and total number of voting lists by $L$, the recursion for computing the trustworthiness $T_r$ of a rater $V_r$ proposed in \cite{lazytom} was given by
\[T_r^{(n+1)}=\frac{1}{L}\sum_{l,i}\left(\frac{\sum_{m\,:\,m,r\rightarrow li}T_m^{(n)}}{\sum_{m=1}^{N}T_m^{(n)}}\right)^{\frac{p}{p+1}}.\]
Unfortunately, the normalizing factor in the denominator on the righthand side can become excessive as the number of voters who did not vote in any elections in which $V_r$ has voted increases, making the rank computation unstable. Also, the exponent $\frac{p}{p+1}$ is always smaller than $1$, and this severely limits the robustness of the proposed method against collusion attacks. The algorithm aimed to relate (a power of) the ratios between trustworthiness of any two voters to the ratios of the numbers of votes received by the candidates chosen by these voters. It also normalized the  trustworthiness of voters, instead of normalizing the credibility of levels; however, as we do it in our present algorithm, normalizing credibility of levels, which are going to be used as weights in a subsequent (independent) computation of ranks of objects, not only makes more sense but also allows an elegant proof of convergence, missing in \cite{lazytom}.

In summary, unlike the existing models for collusion detection, we do NOT rely on any clustering techniques, local indicators or averaging; also, our  method does not rely on any approximation of the final rating scores, making rating an entirely independent process from the credibility assessment.

\section{Conclusion}\label{sec:con}

As we have mentioned, existing iterative methods, such as~\cite{iterativefilter1,iterativefilter2,de1,de2,de3}, approximate ranks using techniques involving weighted averages. However, averages generally have the propensity to blur statistical features because they smooth out data. In our method, the trustworthiness of raters is computed purely from the concurrence of opinions, without any averaging at all. In fact, note that in our ``rating-through-voting'' method, the ordering of the range of credibility levels (i.e., an increasing ordering from e.g., $1$ to $10$) is NOT considered at all - we treat such domain as an unordered set, and only consider the concurrence of opinions. Such obtained trustworthiness of raters and the credibility of items (in this case rating levels) can then be used to obtain the values of the rating scores in a completely decoupled way, for example, by taking a weighted average with weights obtained as some function of the credibility scores obtained for each rating level, or by choosing the highest ranked level or many other possible ways, depending on what kind of statistical feature we are mining the submitted evaluation data.

However, in our future work, we will further refine our method by taking into account the ordering of the rating levels, with the aim to produce a complete yet fully  flexible rating methodology which can be precisely tuned to produce rating scores reflecting any desired  statistical feature of evaluation data.

\balance

\bibliographystyle{abbrv}
\bibliography{VLDB_v3.0.bbl}

\begin{thebibliography}{10}

\bibitem{FIM}
R.~Agrawal and R.~Srikant.
\newblock Fast algorithms for mining association rules in large databases.
\newblock In {\em Proceedings of the 20th International Conference on Very
  Large Data Bases}, VLDB '94, pages 487--499, San Francisco, CA, USA, 1994.
  Morgan Kaufmann Publishers Inc.

\bibitem{aleksmarking}
P.~C. C.~C. Aleks~Ignjatovic, Chung Tong~Lee and H.~Guo.
\newblock Computing marks from multiple assessors using adaptive averaging.
\newblock In {\em ICEE}, 2009.

\bibitem{amazon}
Amazon-Online-Market.
\newblock http://www.amazon.com/.

\bibitem{csproblemsolving}
D.~Brabham.
\newblock Crowdsourcing as a model for problem solving an introduction and
  cases.
\newblock {\em Convergence: The International Journal of Research into New
  Media Technologies}, 14(1):75--90, 2008.

\bibitem{sensor}
C.~Chou, A.~Ignjatovic~and, and W.~Hu.
\newblock Efficient computation of robust average of compressive sensing data
  in wireless sensor networks in the presence of sensor faults.
\newblock {\em Parallel and Distributed Systems, IEEE Transactions on},
  PP(99):1, 2012.

\bibitem{CollusionInP2P}
G.~Ciccarelli and R.~L. Cigno.
\newblock Collusion in peer-to-peer systems.
\newblock {\em Computer Networks}, 55(15):3517 -- 3532, 2011.

\bibitem{de1}
C.~De~Kerchove and P.~Van~Dooren.
\newblock Iterative filtering for a dynamical reputation system.
\newblock {\em Arxiv preprint arXiv:0711.3964}, 2007.

\bibitem{de2}
C.~De~Kerchove and P.~Van~Dooren.
\newblock Reputation systems and optimization.
\newblock {\em Siam News}, 41(2), 2008.

\bibitem{de3}
C.~de~Kerchove and P.~Van~Dooren.
\newblock Iterative filtering in reputation systems.
\newblock {\em SIAM J. Matrix Anal. Appl.}, 31(4):1812--1834, Mar. 2010.

\bibitem{cswww}
A.~Doan, R.~Ramakrishnan, and A.~Y. Halevy.
\newblock Crowdsourcing systems on the world-wide web.
\newblock {\em Commun. ACM}, 54:86--96, April 2011.

\bibitem{glens}
GroupLens-Research-Lab.
\newblock http://www.grouplens.org/.

\bibitem{amazonproblem}
A.~HARMON.
\newblock Amazon glitch unmasks war of reviewers.
\newblock In {\em NY Times (2004, Feb. 14)}.

\bibitem{aleksrep}
A.~Ignjatovic, N.~Foo, and C.~T. Lee.
\newblock An analytic approach to reputation ranking of participants in online
  transactions.
\newblock In {\em Proceedings of the 2008 IEEE/WIC/ACM International Conference
  on Web Intelligence and Intelligent Agent Technology - Volume 01}, pages
  587--590, Washington, DC, USA, 2008. IEEE Computer Society.

\bibitem{IMDBR}
IMDB-Rating-Explanation.
\newblock https://resume.imdb.com/help/show\_leaf?votes.

\bibitem{imdb}
Internet-Movie-DataBase.
\newblock http://www.imdb.com/.

\bibitem{ebayProblem}
J.~M. J.~Brown.
\newblock Reputation in online auctions: The market for trust.
\newblock {\em CALIFORNIA MANAGEMENT REVIEW}, 49(1):61 --81, Fall 2006.

\bibitem{maj2}
W.~Jianshu, M.~Chunyan, and G.~Angela.
\newblock An entropy-based approach to protecting rating systems from unfair
  testimonies.
\newblock {\em IEICE TRANSACTIONS on Information and Systems},
  89(9):2502--2511, 2006.

\bibitem{unusuall}
N.~Jindal, B.~Liu, and E.-P. Lim.
\newblock Finding unusual review patterns using unexpected rules.
\newblock In {\em Proceedings of the 19th ACM international conference on
  Information and knowledge management}, CIKM '10, pages 1549--1552, New York,
  NY, USA, 2010. ACM.

\bibitem{eigentrust}
S.~D. Kamvar, M.~T. Schlosser, and H.~Garcia-Molina.
\newblock The eigentrust algorithm for reputation management in p2p networks.
\newblock In {\em Proceedings of the 12th international conference on World
  Wide Web}, WWW '03, pages 640--651, New York, NY, USA, 2003. ACM.

\bibitem{iterativefilter1}
P.~Laureti, L.~Moret, Y.~Zhang, and Y.~Yu.
\newblock Information filtering via iterative refinement.
\newblock {\em EPL (Europhysics Letters)}, 75:1006, 2006.

\bibitem{lazytom}
C.~T. Lee, E.~M. Rodrigues, G.~Kazai, N.~Milic-Frayling, and A.~Ignjatovic.
\newblock Model for voter scoring and best answer selection in community q\&a
  services.
\newblock {\em Web Intelligence and Intelligent Agent Technology, IEEE/WIC/ACM
  International Conference on}, 1:116--123, 2009.

\bibitem{EigenCollusion}
Q.~Lian, Z.~Zhang, M.~Yang, B.~Y. Zhao, Y.~Dai, and X.~Li.
\newblock An empirical study of collusion behavior in the maze p2p file-sharing
  system.
\newblock In {\em Proceedings of the 27th International Conference on
  Distributed Computing Systems}, ICDCS '07, pages 56--, Washington, DC, USA,
  2007. IEEE Computer Society.

\bibitem{CIKMInds}
E.-P. Lim, V.-A. Nguyen, N.~Jindal, B.~Liu, and H.~W. Lauw.
\newblock Detecting product review spammers using rating behaviors.
\newblock In {\em Proceedings of the 19th ACM international conference on
  Information and knowledge management}, CIKM '10, pages 939--948, New York,
  NY, USA, 2010. ACM.

\bibitem{signaldetection}
Y.~Liu, Y.~Yang, and Y.~Sun.
\newblock Detection of collusion behaviors in online reputation systems.
\newblock In {\em Signals, Systems and Computers, 2008 42nd Asilomar Conference
  on}, pages 1368--1372. IEEE, 2008.

\bibitem{mlens}
MovieLens.
\newblock http://movielens.org/.

\bibitem{dataset}
MovieLens-dataset.
\newblock http://www.grouplens.org/node/73/.

\bibitem{mtrust}
MovieTrust.
\newblock http://www.cse.unsw.edu.au/~mallahbakhsh/mtrust/.

\bibitem{mainWWW}
A.~Mukherjee, B.~Liu, and N.~Glance.
\newblock Spotting fake reviewer groups in consumer reviews.
\newblock In {\em Proceedings of the 21st international conference on World
  Wide Web}, WWW '12, pages 191--200, New York, NY, USA, 2012. ACM.

\bibitem{pagerank}
L.~Page, S.~Brin, R.~Motwani, and T.~Winograd.
\newblock The pagerank citation ranking: Bringing order to the web.
\newblock Technical Report 1999-66, Stanford InfoLab, November 1999.
\newblock Previous number = SIDL-WP-1999-0120.

\bibitem{rt}
Rotten-Tomatoes.
\newblock http://www.rottentomatoes.com/.

\bibitem{so}
StackOverflow.
\newblock http://stackoverflow.com.

\bibitem{IEEE2102Survey}
Y.~Sun and Y.~Liu.
\newblock Security of online reputation systems: The evolution of attacks and
  defenses.
\newblock {\em Signal Processing Magazine, IEEE}, 29(2):87 --97, march 2012.

\bibitem{wisdom}
J.~Surowiecki.
\newblock {\em The Wisdom of Crowds}.
\newblock Anchor Books, 2005.

\bibitem{reliable}
G.~Swamynathan, K.~Almeroth, and B.~Zhao.
\newblock The design of a reliable reputation system.
\newblock {\em Electronic Commerce Research}, 10:239--270, 2010.
\newblock 10.1007/s10660-010-9064-y.

\bibitem{ESP}
L.~von Ahn and L.~Dabbish.
\newblock Designing games with a purpose.
\newblock {\em Commun. ACM}, 51:58--67, August 2008.

\bibitem{spam}
J.~Vuurens and A.~de~Vries.
\newblock Obtaining high-quality relevance judgments using crowdsourcing.
\newblock {\em Internet Computing, IEEE}, PP(99):1, 2012.

\bibitem{turf}
G.~Wang, C.~Wilson, X.~Zhao, Y.~Zhu, M.~Mohanlal, H.~Zheng, and B.~Y. Zhao.
\newblock Serf and turf: crowdturfing for fun and profit.
\newblock In {\em Proceedings of the 21st international conference on World
  Wide Web}, WWW '12, pages 679--688, New York, NY, USA, 2012. ACM.

\bibitem{reptrap}
Y.~Yang, Q.~Feng, Y.~L. Sun, and Y.~Dai.
\newblock Reptrap: a novel attack on feedback-based reputation systems.
\newblock In {\em Proceedings of the 4th international conference on Security
  and privacy in communication netowrks}, SecureComm '08, pages 8:1--8:11, New
  York, NY, USA, 2008. ACM.

\bibitem{dishonest}
Y.-F. Yang, Q.-Y. Feng, Y.~Sun, and Y.-F. Dai.
\newblock Dishonest behaviors in online rating systems: cyber competition,
  attack models, and attack generator.
\newblock {\em J. Comput. Sci. Technol.}, 24(5):855--867, Sept. 2009.

\bibitem{yelp}
Yelp.
\newblock http://www.yelp.com.

\bibitem{iterativefilter2}
Y.-K. Yu, Y.-C. Zhang, P.~Laureti, and L.~Moret.
\newblock Decoding information from noisy, redundant, and intentionally
  distorted sources.
\newblock {\em Physica A: Statistical Mechanics and its Applications},
  371(2):732 -- 744, 2006.

\bibitem{HCOmpsurvey}
M.-C. Yuen, L.-J. Chen, and I.~King.
\newblock A survey of human computation systems.
\newblock In {\em Computational Science and Engineering, 2009. CSE '09.
  International Conference on}, volume~4, pages 723 --728, aug. 2009.

\end{thebibliography}

\end{document}